\documentstyle[onecolumn,epsfig]{mn}

\newif\ifAMStwofonts
\def \half{\frac{1}{2}}
\def \eps{\epsilon}
\def \d{\partial}
\def \lam{\lambda}
\def \non{\nonumber}
\def \cLa{{\cal L}_1^{\pm 1}}
\def \cLb{{\cal L}_2^{\pm 1}}
\def \cLc{{\cal L}_3^{\pm 1}}
\def \cLd{{\cal L}_4^{\pm 1}}
\renewcommand{\(}{\left(}
\renewcommand{\)}{\right)}
\newcommand{\df}[2]{ \frac{\partial {#1}}{\partial {#2}} }
\newcommand{\dff}[2]{ \frac{\partial^2 {#1}}{\partial {#2}^2} }

\newcommand{\I}{\mbox{\rm i}}
\newcommand{\beq}{\begin{equation}}
\newcommand{\eeq}{\end{equation}}
\newcommand{\beqar}{\begin{eqnarray}}
\newcommand{\eeqar}{\end{eqnarray}}
\newcommand{\beqars}{\begin{eqnarray*}}
\newcommand{\eeqars}{\end{eqnarray*}}
\newcommand{\bc}{\begin{center}}
\newcommand{\ec}{\end{center}}
\newcommand{\ben}{\begin{enumerate}}
\newcommand{\een}{\end{enumerate}}
\newcommand{\bit}{\begin{itemize}}
\newcommand{\eit}{\end{itemize}}

\ifoldfss
  \ifCUPmtlplainloaded \else
    \NewTextAlphabet{textbfit} {cmbxti10} {}
    \NewTextAlphabet{textbfss} {cmssbx10} {}
    \NewMathAlphabet{mathbfit} {cmbxti10} {} % for math mode
    \NewMathAlphabet{mathbfss} {cmssbx10} {} %  "   "    "
  \fi
  \ifAMStwofonts
    \ifCUPmtlplainloaded \else
      \NewSymbolFont{upmath} {eurm10}
      \NewSymbolFont{AMSa} {msam10}
      \NewMathSymbol{\upi}     {0}{upmath}{19}
      \NewMathSymbol{\umu}     {0}{upmath}{16}
      \NewMathSymbol{\upartial}{0}{upmath}{40}
      \NewMathSymbol{\leqslant}{3}{AMSa}{36}
      \NewMathSymbol{\geqslant}{3}{AMSa}{3E}

    \fi
  \fi
\fi % End of OFSS

\ifnfssone
  \newmathalphabet{\mathit}
  \addtoversion{normal}{\mathit}{cmr}{m}{it}
  \addtoversion{bold}{\mathit}{cmr}{bx}{it}
  \newmathalphabet{\mathbfit} % math mode version of \textbfit{..}
  \addtoversion{normal}{\mathbfit}{cmr}{bx}{it}
  \addtoversion{bold}{\mathbfit}{cmr}{bx}{it}
  \newmathalphabet{\mathbfss} % math mode version of \textbfss{..}
  \addtoversion{normal}{\mathbfss}{cmss}{bx}{n}
  \addtoversion{bold}{\mathbfss}{cmss}{bx}{n}
  \ifAMStwofonts
    \ifCUPmtlplainloaded \else
      \UseAMStwoboldmath
      \makeatletter
      \new@mathgroup\upmath@group
      \define@mathgroup\mv@normal\upmath@group{eur}{m}{n}
      \define@mathgroup\mv@bold\upmath@group{eur}{b}{n}
      \edef\UPM{\hexnumber\upmath@group}
      \new@mathgroup\amsa@group
      \define@mathgroup\mv@normal\amsa@group{msa}{m}{n}
      \define@mathgroup\mv@bold\amsa@group{msa}{m}{n}
      \edef\AMSa{\hexnumber\amsa@group}
      \makeatother
      \mathchardef\upi="0\UPM19
      \mathchardef\umu="0\UPM16
      \mathchardef\upartial="0\UPM40
      \mathchardef\leqslant="3\AMSa36
      \mathchardef\geqslant="3\AMSa3E
    \fi
  \fi
\fi % End of NFSS release 1

\ifnfsstwo
  \DeclareMathAlphabet{\mathbfit}{OT1}{cmr}{bx}{it}
  \SetMathAlphabet\mathbfit{bold}{OT1}{cmr}{bx}{it}
  \DeclareMathAlphabet{\mathbfss}{OT1}{cmss}{bx}{n}
  \SetMathAlphabet\mathbfss{bold}{OT1}{cmss}{bx}{n}
  \ifAMStwofonts
    \ifCUPmtlplainloaded \else
      \DeclareSymbolFont{UPM}{U}{eur}{m}{n}
      \SetSymbolFont{UPM}{bold}{U}{eur}{b}{n}
      \DeclareSymbolFont{AMSa}{U}{msa}{m}{n}
      \DeclareMathSymbol{\upi}{0}{UPM}{"19}
      \DeclareMathSymbol{\umu}{0}{UPM}{"16}
      \DeclareMathSymbol{\upartial}{0}{UPM}{"40}
      \DeclareMathSymbol{\leqslant}{3}{AMSa}{"36}
      \DeclareMathSymbol{\geqslant}{3}{AMSa}{"3E}
    \fi
  \fi
\fi % End of NFSS release 2

\ifCUPmtlplainloaded \else
  \ifAMStwofonts \else % If no AMS fonts
    \def\upi{\pi}
    \def\umu{\mu}
    \def\upartial{\partial}
  \fi
\fi

\title[Evolution equations for the perturbations of slowly rotating
relativistic stars]{Evolution equations for the perturbations
  of slowly rotating relativistic stars}

\author[J.~Ruoff, A.~Stavridis and K.D.~Kokkotas]
{Johannes Ruoff, Adamantios Stavridis and Kostas D.~Kokkotas\\
  Department of Physics, Aristotle University of Thessaloniki,
  Thessaloniki 54006, Greece}

\date{Accepted ???? Month ??.
      Received ???? Month ??;
      in original form ???? Month ??}

\pagerange{\pageref{firstpage}--\pageref{lastpage}}
\pubyear{2001}
\begin{document}

\maketitle

\label{firstpage}

\begin{abstract}
  We present a new derivation of the equations governing the
  oscillations of slowly rotating relativistic stars. Previous
  investigations have been mostly carried out in the Regge--Wheeler
  gauge. However, in this gauge the process of linearizing the
  Einstein field equations leads to perturbation equations which as
  such cannot be used to perform numerical time evolutions. It is only
  through the tedious process of combining and rearranging the
  perturbation variables in a clever way that the system can be cast
  into a set of hyperbolic first order equations, which is then well
  suited for the numerical integration. The equations remain quite
  lengthy, and we therefore rederive the perturbation equations in a
  different gauge, which has been first proposed by Battiston et
  al.~(1970). Using the ADM formalism, one is immediately lead to a
  first order hyperbolic evolution system, which is remarkably simple
  and can be numerically integrated without many further
  manipulations. Moreover, the symmetry between the polar and the
  axial equations becomes directly apparent.
\end{abstract}


\section{Introduction}

The theory of non-radial perturbations of relativistic stars has been
a field of intensive study for more than three decades, beginning with
the pioneering paper of Thorne \& Campolattaro in 1967. Their work was
in turn based on previous studies of black hole perturbations
initiated by Regge \& Wheeler in 1957. Because of the quite involved
and tedious computations of the perturbed field equations, the main
focus remained on non-rotating neutron stars, although the foundations
for computing rotating relativistic stellar models have already been
laid by Hartle in 1967.  Only in the early 90s, work began to shift to
the perturbations of rotating relativistic neutron stars. In most
works, the slow-rotation approximation was used to tackle the problem.
Chandrasekhar \& Ferrari (1991) studied the axisymmetric
perturbations, where they established the coupling between the polar
and axial modes induced by the rotation (polar or even parity modes
are characterised by a sign change under parity transformation
according to $(-1)^l$ while the axial ones change as $(-1)^{l+1}$).
Shortly later, Kojima (1992) presented the first complete derivation
of the coupled polar and axial perturbation equations.  These
equations were the starting point for investigations of relativistic
rotational effects on stellar oscillations and associated
instabilities. Most of the work considered the simpler task of solving
the perturbation equations in the frequency domain and, as a result,
calculations involved the determination of the eigenfrequencies rather
than the solution of the time dependent equations. A somewhat
different approach based on a Lagrangian description was used by
Lockitch et al.~(2001) with the focus on the computation of
rotationally induced inertial modes.

Following the more than 40 years old tradition, it was quite common to
work in the Regge--Wheeler gauge, although some groups were using
different gauges or the gauge invariant formulation of Moncrief
(1974). In particular, a gauge introduced by Battiston, Cazzola \&
Lucaroni in 1971 in a series of papers to study the stability
properties of non-radial oscillations in relativistic non-rotating
stars, has not received proper attention (Battiston, Cazzola \&
Lucaroni 1971, Cazzola \& Lucaroni 1972, 1974, 1978; Cazzola, Lucaroni
\& Semezato 1978).

Since the perturbation equations of non-rotating stars are fairly
simple, there is no real advantage of one gauge over the other.  Note,
however, that there was a long standing puzzle why in the
Regge--Wheeler gauge the equations could be reduced to a fourth order
system, whereas in the diagonal gauge used by Chandrasekhar \& Ferrari
(1991), which was previously used by Chandrasekhar (1983) in the
context of black hole perturbations, only a fifth order system could
be obtained. The latter system gave rise to an additional divergent
solution, which had to be rejected on physical grounds.  This
discrepancy was finally solved by Price \& Ipser (1991), who showed
that the diagonal gauge was not completely fixed and still possessed
one degree of gauge freedom, giving rise to the additional spurious
solution.

As computer power has been enormously increasing within the last
decade, the problem of evolving the fully non-linear 3D Einstein
equations in the time domain has finally become into reach of
feasibility.  Various groups around the world are now building robust
codes that perform the time evolution of both single neutron stars or
binaries in full 3D (see for instance Font et al.~2001,
Shibata \& Uryu 2001, and the review of Stergioulas 1998).
Nevertheless, there is still considerable work in progress within
perturbative approaches, which can provide us with a deeper
understanding of the physics of relativistic star perturbations.
Moreover, any trustable non-linear code must be able to reproduce the
results from perturbation theory.

Time evolutions of the perturbative equations have been carried out
first for the axial equations (Andersson \& Kokkotas 1996) and then
for the polar equations using the Regge--Wheeler gauge by Allen et
al.~(1998) and Ruoff (2001a).  Allen et al.~(1998) managed to write
down the evolution equations as two relatively simple wave equations
for the metric perturbations and one wave equation for the fluid
enthalpy perturbation inside the star. Ruoff (2001a) rederived these
equations using the ADM formalism (Arnowitt et al.~1962). They were
used to evolve and study initial data, representing the late stage of
a binary neutron star head-on collision (Allen et al.~1999).

Using the Arnowitt--Deser--Misner (ADM) formalism, the evolution
equations for the axial perturbations of rotating stars have been
brought into a suitable form for the numerical evolution by Ruoff \&
Kokkotas (2001a,b). Here, the resulting system of equations came out
immediately as a first order system both in space and time, and it
could be directly used for the numerical evolution without many
further manipulations. In the non-rotating case, it is an easy task to
transform the first order system into a single wave equation for just
one metric variable. In the rotating case, however, this is not
possible any more because of the rotational correction terms.

When looking at the set of polar equations derived by Kojima (1992) it
appears clearly that the presence of mixed spatial and time
derivatives makes them not suitable for the numerical time
integration.  Nevertheless, using a number of successive manipulations
and the introduction of many additional variables, we were able to
cast the equations into a hyperbolic first order form.

A more natural way to automatically obtain a set of equations, which
is first order in time, is to use the ADM formalism. However, as we
shall explain, even in that case the polar equations in the
Regge--Wheeler gauge need to be further manipulated before they are
suitable for a numerical integration. In general the ADM formalism
yields a set of partial differential equations which are first order
in time, but second order in space.  For the numerical evolution, this
is not ideal and one would rather prefer to have a pure first order
system, or if possible a pure second order system, thus representing
generalized wave equations. As we mentioned above, in the non-rotating
case, it is easily possible to convert the perturbation equations into
a set of wave equations.  However, in the rotating case, this is not
possible any more, even in the simple case when only axial
perturbations are considered. To illustrate the problems associated
with the Regge-Wheeler gauge, let us recall Einstein's (unperturbed)
evolution equations written in the ADM formalism:
\begin{eqnarray}\label{eq_g}
  \(\d_t - \cal L_\beta\)\gamma_{ij} &=& - 2\alpha K_{ij}\;,\\
  \label{eq_K}
  \(\d_t - \cal L_\beta\) K_{ij} &=& - \alpha_{;ij}
  + \alpha \left[R_{ij} + K^k_{\phantom{k}k}K_{ij} - 2K_{ik}K^k_{\phantom{i}j} - 4\pi
    \(2T_{ij} - T^\nu_{\phantom{\nu}\nu}\gamma_{ij}\) \right]\;,
\end{eqnarray}
with $\alpha$ denoting the lapse function, $\beta^i$ the shift vector,
$\cal L_\beta$ the Lie-derivative with respect to $\beta^i$,
$\gamma_{ij}$ the metric of a space-like three dimensional
hypersurface with Ricci tensor $R_{ij}$, and $K_{ij}$ its extrinsic
curvature.  It is obvious that the only second order spatial
derivatives are $\d_i\d_j\alpha$ and $\d_i\d_j\gamma_{kl}$ with the
latter coming from the Ricci tensor $R_{ij}$. This is still true for
the linearized version of Eqs.~(\ref{eq_g}) and (\ref{eq_K}).

In the Regge--Wheeler gauge, we have a non-vanishing perturbation of
the lapse $\alpha$ and of the diagonal components of the spatial
perturbations $h_{ij}$. Using the notation of Ruoff (2001a), the polar
perturbations can be written as
\begin{eqnarray}
    \alpha &\sim& \sum_{l,m}{S_1^{lm}(t,r)Y_{lm}(\theta,\phi)}\;,\\
    h_{ij} &\sim&   \sum_{l,m}{\(
  \begin{array}{ccc}
    S_3^{lm}(t,r) & 0 & 0\\
    0 & T_2^{lm}(t,r) & 0\\
    0 & 0 & \sin^2\theta\,T_2^{lm}(t,r)\\
  \end{array}\)Y_{lm}(\theta,\phi)}\;.
\end{eqnarray}
The perturbation equations coming from Eq.~(\ref{eq_K}) contain second
order $r$-derivatives of $S_1$ and $T_2$. Note that they do not
contain second order derivatives of $S_3$, because only the angular
components of the metric get differentiated twice with respect to $r$.
In the axial case there is only one perturbation function for the
angular metric components, but it is set to zero in the Regge--Wheeler
gauge. Therefore the ADM formalism immediately yields a first order
system.

The polar equations, in contrast, can be cast into a fully first order
system only through the introduction of some auxiliary variables. In
the non-rotating case, this is a fairly easy task, but for the
rotating case, it turns out to become considerably more complicated.
This is because the simple proportionality of $S_1$ and $S_3$, which
holds in the non-rotating case and which makes is easy to replace
$S_1$ by $S_3$, does not do so any more in the rotating case. Instead
this relation involves rotational correction terms, and the
replacement of $S_1$ by $S_3$ would lead to a considerable inflation
of the equations.

Consequently, instead of manipulating the perturbation equations in
the Regge--Wheeler gauge, we look for a gauge in which the
perturbation equations, by construction, do not show any second order
spatial derivative. We have seen that the second derivatives come from
the angular terms in the spatial metric and the lapse function. It
seems therefore natural to set these components to zero. For the axial
case this is already realized in the Regge--Wheeler gauge. It is only
for the polar perturbations that we need a different gauge. From the
seven polar components of the metric, Regge \& Wheeler (1957) chose to
set the components $V_1$, $V_3$ and $T_1$, which represent in the
notation of Ruoff (2001a) the polar angular vector perturbations and
one of the angular tensor perturbations, to zero. We now choose a
different set, namely we set the angular terms in the spatial metric
$T_1$, $T_2$ together with the lapse $S_1$ to zero and retain $V_1$
and $V_3$. With this choice we expect the ADM formalism to provide us
with an evolution system without any second $r$-derivatives.

We should mention again that this gauge has actually been introduced
thirty years ago by Battiston, Cazzola \& Lucaroni (1971) to derive
the perturbation equations for non-radial oscillations of non-rotating
neutron stars and to investigate in a subsequent series of papers
their stability properties (Cazzola \& Lucaroni 1972, 1974, 1978;
Cazzola, Lucaroni \& Semezato 1978a, 1978b). Particularly relevant is
the first paper of the series, in which they proved the uniqueness of
this gauge, hereafter referred to as the BCL gauge, and where they
also showed the relation with the Regge--Wheeler gauge.

In Section 2, we will use the ADM formalism to derive the time
dependent perturbation equations for slowly rotating relativistic
stars in the BCL gauge. Section 3 contains a brief discussion of the
non-rotating limit and conclusions will be given in Section 4. In the
Appendix, we present the perturbation equations as they follow
directly from Einstein's equations in a form similar to the equations
in the Regge--Wheeler gauge given by Kojima (1992). Throughout the
paper, we adopt the metric signature $(- + + +)$, and we use
geometrical units with $c=G=1$.  Derivatives with respect to the
radial coordinate $r$ are denoted by a prime, while derivatives with
respect to time $t$ are denoted by an over-dot. Greek indices run from
0 to 3, Latin indices from 1 to 3.

\section{The perturbation equations in the ADM formalism}

The metric describing a slowly rotating neutron star reads in spherical
coordinates ($t$, $r$, $\theta$, $\phi$)
\begin{eqnarray}\label{metric}
  g_{\mu\nu} &=& \(
  \begin{array}{cccc}
    -e^{2\nu} & 0 & 0 & -\omega r^2 \sin^2\theta\\
    0 & e^{2\lam} & 0 & 0\\
    0 & 0 & r^2 & 0\\
    -\omega r^2 \sin^2\theta & 0 & 0 & r^2\sin^2\theta\\
  \end{array}\)\;,
\end{eqnarray}
where $\nu$, $\lam$ and the ``frame dragging'' $\omega$ are
functions of the radial coordinate $r$ only. With the neutron star matter
described by a perfect fluid with pressure $p$, energy density
$\eps$, and 4-velocity
\begin{equation}
  U^\mu = \(e^{-\nu}, 0, 0, \Omega e^{-\nu}\)\;,
\end{equation}
the Einstein equations together with an equation of state $p=p(\eps)$
yield the well known TOV equations plus an extra equation for the
frame dragging. To linear order, it is given by
\begin{equation}\label{drag}
  \varpi'' - \(4\pi re^{2\lam}(p + \eps) - \frac{4}{r}\)\varpi'
  - 16\pi e^{2\lam}\(p + \eps\)\varpi = 0\;,
\end{equation}
where
\begin{equation}
  \varpi := \Omega - \omega
\end{equation}
represents the angular velocity of the fluid relative to the local
inertial frame. In the language of the ADM formalism, we have to
express the above background metric (\ref{metric}) in terms of lapse ,
(covariant) shift and the 3-metric, which we denote by $A$, $B_i$ and
$\gamma_{ij}$, respectively.  Explicitly, we have
\begin{eqnarray}
  A &=& \sqrt{B^iB_i - g_{00}} \;=\; e^{\nu} + O(\omega^2)\;,\\
  B_i &=& \(0, 0, -\omega r^2 \sin^2\theta\)\;,\\
  \gamma_{ij} &=& \(
  \begin{array}{ccc}
    e^{2\lam} & 0 & 0\\
    0 & r^2 & 0\\
    0 & 0 & r^2\sin^2\theta\\
  \end{array}\)\;.
\end{eqnarray}
The extrinsic curvature of the space-like hypersurface described by
$\gamma_{ij}$ can be computed by
\begin{eqnarray}
  K_{ij} &=& \frac{1}{2A}\(B^k\d_k\gamma_{ij}
  + \gamma_{ki}\d_jB^k + \gamma_{kj}\d_iB^k\)\;,
\end{eqnarray}
yielding the only non-vanishing components
\begin{eqnarray}
  K_{13} &=& K_{31} \;=\; -\half\omega' e^{-\nu} r^2\sin^2\theta\;.
\end{eqnarray}
The perturbations of the background lapse $A$, shift $B_i$, 3-metric
$\gamma_{ij}$, extrinsic curvature $K_{ij}$, 4-velocity $U_i$, energy
density $\eps$ and pressure $p$ will be denoted by $\alpha$,
$\beta_i$, $h_{ij}$, $k_{ij}$, $u_i$, $\delta\eps$ and $\delta p$,
respectively. The twelve evolution equations for $h_{ij}$ and $k_{ij}$
are obtained by linearizing the non-linear ADM equations (\ref{eq_g})
and (\ref{eq_K}). Working in the slow-rotation approximation, we keep
only terms up to order $\Omega$ (or $\omega$). The background
quantities $B^k$ and $K_{ij}$ are first order in $\Omega$, hence we
can neglect any products thereof. Furthermore, it is $K =
\gamma^{ij}K_{ij} = 0$.  These circumstances lead to cancellations of
various terms and the perturbations equations reduce to:
\begin{eqnarray}
  \label{dtgij}
  \d_t h_{ij} &=& \d_i\beta_j + \d_j\beta_i
  - 2\(A k_{ij} + K_{ij}\alpha
  + \Gamma^k_{\phantom{k}ij}\beta_k +
  B_k \delta\Gamma^k_{\phantom{k}ij}\)\;,\\
\label{dtKij}
  \d_t k_{ij} &=& \alpha \left[R_{ij} + 4\pi(p - \eps)\gamma_{ij}\right]
  - \d_i\d_j\alpha + \Gamma^k_{\phantom{i}ij}\d _k\alpha
  + \delta \Gamma^k_{\phantom{i}ij}\d_kA\non\\
  &&+ A \left[\delta R_{ij}
    + K_{ij}k
    - 2\(K_i^{\phantom{i}k}k_{jk} + K_j^{\phantom{i}k}k_{ik}\)
    + 4\pi\((p - \eps)h_{ij}
    + \gamma_{ij}\(\delta p - \delta\eps\)
    - 2(p + \eps)\(u_i\delta u_j + u_j\delta u_i\)\)\right] \non\\
  && + B^k\d_kk_{ij}
  + \(\d_kK_{ij}
  - K_i^{\phantom{i}l}\d_j\gamma_{kl}
  - K_j^{\phantom{i}l}\d_i\gamma_{kl}\)\beta^k
  + k_{ik}\d_jB^k + k_{jk}\d_iB^k
  + K_i^{\phantom{i}k}\d_j\beta_k
  + K_j^{\phantom{i}k}\d_i\beta_k\;,
\end{eqnarray}
where
\begin{eqnarray}
  k &:=& \gamma^{ij}k_{ij}\;,\\
  \delta\Gamma^k_{\phantom{k}ij} &:=&
  \half\gamma^{km}\(\d_ih_{mj} + \d_jh_{mi} - \d_mh_{ij}
  - 2\Gamma^{l}_{\phantom{i}ij} h_{lm}\)\;,\\
  \delta R_{ij} &:=& \d_k\delta\Gamma^k_{\phantom{i}ij}
  - \d_j\delta\Gamma^k_{\phantom{i}ik}
  + \Gamma^l_{\phantom{i}ij}\delta\Gamma^k_{\phantom{i}lk}
  + \Gamma^k_{\phantom{i}lk}\delta\Gamma^l_{\phantom{i}ij}
  - \Gamma^l_{\phantom{i}ik}\delta\Gamma^k_{\phantom{i}lj}
  - \Gamma^k_{\phantom{i}lj}\delta\Gamma^l_{\phantom{i}ik}\;.
\end{eqnarray}
To obtain a closed set of evolution equations, we will also use the
four evolution equations the fluid perturbations following from the
linearized conservation law $\delta T^{\mu\nu}_{\phantom{\mu\nu};\mu}
= 0$. Last but not least we need the four linearized constraint
equations, which serve to construct physically valid initial data and
to monitor the accuracy of the numerical evolution:
\begin{eqnarray}\label{hhc}
  \gamma^{ij}\delta R_{ij} - h^{ij}R_{ij} - 2K^{ij}k_{ij}
  &=& 16\pi\(\delta\eps + 2e^{-\nu}(p + \eps)(\Omega - \omega)\delta 
  u_3\)\;,\\
  \label{mmc}
  -8\pi\left[(p + \eps)\delta u_i + u_i\(\delta p + \delta\eps\)\right]&=&
  \gamma^{jk}\(\d _i k_{jk} - \d _j k_{ik}
  - \Gamma^l_{\phantom{i}ik}k_{jl} + \Gamma^l_{\phantom{i}jk} k_{il}
  - \delta\Gamma^l_{\phantom{i}ik}K_{jl}
  + \delta\Gamma^l_{\phantom{i}jk}K_{il}\)\non\\
  &&- h^{jk}\(\d _i K_{jk} - \d _j K_{ik}
  - \Gamma^l_{\phantom{i}ik}K_{jl} + \Gamma^l_{\phantom{i}jk}K_{il}\)\;.
\end{eqnarray}
We assume the oscillations to be adiabatic, thus the relation between
the Eulerian pressure perturbation $\delta p$ and density perturbation
$\delta\eps$ is given by
\begin{equation}
  \label{adcond}
  \delta p = \frac{\Gamma_1p}{p + \eps}\delta\eps
  + p'\xi^r\(\frac{\Gamma_1}{\Gamma} - 1\)\;,
\end{equation}
where $\Gamma_1$ represents the adiabatic index of the perturbed
configuration, $\Gamma$ is the background adiabatic index
\begin{equation}
  \Gamma = \frac{p + \eps}{p}\frac{dp}{d\eps}\;,
\end{equation}
and $\xi^r$ is the radial component of the fluid displacement vector
$\xi^\mu$. The latter is related to the (covariant) 4-velocity
perturbations $\delta u_\mu$ as follows
\begin{equation}
  \delta u_\mu = u^\nu h_{\mu\nu} + g_{\mu\nu}u^\lam\frac{\d\xi^\nu}{\d x^\lam}
  - \half u_\mu u^\kappa u^\lam h_{\kappa\lam}\;.
\end{equation}
For the $r$ component, this gives us
\begin{equation}
  \label{xir}
  \(\d_t + \Omega\d_\phi\)\xi^r =
  e^{-2\lam}\(e^\nu\delta u_r - \beta_r - \Omega h_{r\phi}\)\;.
\end{equation}
To proceed further, we expand the complete set of perturbations
variables into spherical harmonics $Y_{lm} = Y_{lm}(\theta,\phi)$.
This will enable us to eliminate the angular dependence and obtain a
set of equations for the coefficients, which now only depend on $t$
and $r$. It is only then that we can finally choose our gauge. In
principle, choosing the gauge amounts to providing prescriptions for
lapse $\alpha$ and shift $\beta_i$.  Those so-called slicing
conditions determine how the space-like 3-metric foliates the
4-dimensional spacetime. In perturbation theory, the gauge can be used
to set some of the ten metric perturbations to zero. We could, for
instance, set $\alpha = \beta_i = 0$, and we would be left with only
the six components $h_{ij}$. Note, that setting $\alpha$ to zero is
possible, since this is only the perturbation of the background lapse
$A$, and the latter does not vanish.

However, our actual goal is to set some of the spatial perturbation
components $h_{ij}$ to zero, in particular the angular components
$h_{ab}$ with $a,b = \{\theta,\phi\}$. In principle we can prescribe
the values of $h_{ij}$ only once for the initial data, and not
throughout the evolution. The only possible way to keep $h_{ab}$ zero
throughout the evolution is to choose our gauge such that the
evolution equations for $h_{ab}$ become trivial, i.e.~we have to have
\begin{eqnarray}
  \d_t h_{ab} &=& 0\;,\quad a,b = \{\theta,\phi\}\;.
\end{eqnarray}
We will see that this requirement leads to a unique algebraic
condition for the shift vector $\beta_i$. With the definitions
\begin{eqnarray}
  X_{lm} &:=& 2\(\d_\theta - \cot\theta\)\d_\phi Y_{lm}\;,\\
  W_{lm} &:=& \(\d^2_{\theta\theta} - \cot\theta\d_\theta
  - \frac{\d^2_{\phi\phi}}{\sin^2\theta}\)Y_{lm}
  \;=\; \(2\d^2_{\theta\theta} + l(l+1)\)Y_{lm}\;,
\end{eqnarray}
we now expand the metric as follows. For the polar part we choose
(symmetric components are denoted by an asterisk)
\begin{eqnarray}
  \alpha &=& 0\;,\\
  \beta_i^{polar} &=& \sum_{l,m}{\(e^{2\lam}S_2^{lm},\,
    V_1^{lm}\d_\theta,\,V_1^{lm}\d_\phi\)Y_{lm}}\;,\\
  h_{ij}^{polar} &=& \sum_{l,m}{\(\begin{array}{ccc}
    e^{2\lam}S_3^{lm} & V_3^{lm}\d_\theta & V_3^{lm}\d_\phi\\
    \star & 0 & 0\\
    \star & 0 & 0
  \end{array}\)Y_{lm}}\;,\label{p_metric}
\end{eqnarray}
and the axial part is
\begin{eqnarray}
  \beta_i^{axial} &=& \sum_{l,m}{\(0, -V_2^{lm}\frac{\d_\phi}{\sin\theta},\,
    V_2^{lm}\sin\theta\d_\theta\) Y_{lm}}\;,\\
  h_{ij}^{axial} &=& \sum_{l,m}{\(\begin{array}{ccc}
    0 & \displaystyle{-V_4^{lm}\frac{\d_\phi}{\sin\theta}}
    & V_4^{lm} \sin\theta\d_\theta\\
    \star & 0 & 0\\
    \star & 0 & 0
  \end{array}\)Y_{lm}}\;.
\end{eqnarray}
For the extrinsic curvature we have all six components
\begin{eqnarray*}
  k_{ij}^{polar} &=& \half e^{-\nu}\times
\end{eqnarray*}
\begin{equation}
\quad\sum_{l,m}{\(
  \begin{array}{ccc}
    e^{2\lam}K_1^{lm}Y_{lm} & e^{2\lam}K_2^{lm}\d_\theta Y_{lm}&
    e^{2\lam}K_2^{lm}\d_\phi Y_{lm}\\
    \star & \(rK_4^{lm} - \Lambda K_5^{lm}\)Y_{lm} + K_5^{lm} W_{lm}
    & K_5^{lm} X_{lm}\\
    \star & K_5^{lm} X_{lm} &
    \sin^2\theta\left[\(rK_4^{lm} - \Lambda K_5^{lm}\)Y_{lm}
      - K_5^{lm} W_{lm}\right]
  \end{array}\)}\;,
\end{equation}
\begin{eqnarray}
  k_{ij}^{axial} &=& \half e^{-\nu}\sum_{l,m}{
  \(\begin{array}{ccc}
    0 & \displaystyle{ -e^{2\lam}K_3^{lm}\frac{\d_\phi Y_{lm}}{\sin\theta}}
    & e^{2\lam}K_3^{lm}\sin\theta\d_\theta Y_{lm}\\
    \star & \displaystyle{-K_6^{lm}\frac{X_{lm}}{\sin\theta}} & K_6^{lm}\sin\theta\,W_{lm}\\
    \star & K_6^{lm}\sin\theta\,W_{lm} & K_6^{lm}\sin\theta\,X_{lm}
  \end{array}\)}\;.
\end{eqnarray}
Herein and throughout the whole paper, we use the shorthand notation
\begin{equation}
  \Lambda := l(l+1)\;.
\end{equation}
We should note that the somewhat peculiar looking expansions for the
coefficient $K_5^{lm}$ can actually be written as
\begin{eqnarray}
  W_{lm} - \Lambda Y_{lm} &=& 2\d^2_{\theta\theta}Y_{lm}\;,\\
  -\sin^2\theta\(W_{lm} + \Lambda Y_{lm}\) &=&
  2\(\cos\theta\sin\theta\d_\theta + \d^2_{\phi\phi}\)Y_{lm}\;,
\end{eqnarray}
which are essentially the diagonal terms of the Regge--Wheeler tensor
harmonic $\Psi_{\alpha\beta}^{lm}$ (c.f.~Eq.~(20) of Ruoff 2001a).
However, we prefer to write them in terms of $W_{lm}$ and $Y_{lm}$
because it is only for these quantities that simple orthogonality
relations apply. Furthermore, we have to mention that in the
definition of the polar components of the extrinsic curvature, we
differ from the notation of Ruoff (2001a), where the meaning of $K_4$
and $K_5$ is reversed (c.f.~Eq.~(24)). Also, the expansion for the
axial perturbations is not exactly the same as in Ruoff \& Kokkotas
(2001a,b).

In their original paper, Battiston et al.~(1971) did not use the
ADM formalism to fix the gauge, instead they defined their gauge by
directly setting $h_{tt}$, $h_{\theta\theta}$, $h_{\theta\phi}$ and
$h_{\phi\phi}$ to zero. Since the relation between $h_{tt}$ and the
lapse $\alpha$ is given by
\begin{equation}\label{ah_rel}
  h_{tt} = 2A\alpha + 2B^i\beta_i = 2e^{\nu}\alpha - 2\omega h_{t\phi}\;,
\end{equation}
it follows that in the rotating case, $h_{tt} \ne 0$ although the
lapse $\alpha$ vanishes. In the non-rotating case $\beta_i = 0$ and
both $\alpha$ and $h_{tt}$ vanish. If we insisted on keeping a
vanishing $h_{tt}$ also in the rotating case, we would obtain a
non-vanishing lapse, giving us undesired second order spatial
derivatives in the perturbation equations.

Finally the fluid perturbations are decomposed as
\begin{eqnarray}
  \delta u_i^{polar} &=& -e^\nu \sum_{l,m}{
    \(u_1^{lm}, u_2^{lm}\d_\theta, u_2^{lm}\d_\phi\)Y_{lm}}\;,\\
  \delta u_i^{axial} &=& -e^\nu \sum_{l,m}{
    \(0, -u_3^{lm}\frac{\d_\phi}{\sin\theta},
    u_3^{lm}\sin\theta\d_\theta\)Y_{lm}}\;,\\
  \delta \eps &=& \sum_{l,m}{\rho^{lm} Y_{lm}}\;,\\
  \delta p &=& \(p + \eps\)\sum_{l,m}{H^{lm} Y_{lm}}\;,\\
  \xi^r &=& \bigg[\nu'\(1 - \frac{\Gamma_1}{\Gamma}\)\bigg]^{-1}
  \sum_{l,m}{\xi^{lm} Y_{lm}}\;.
\end{eqnarray}
From Eq.~(\ref{adcond}), we have the relation
\begin{equation}
  \rho^{lm} = \frac{\(p + \eps\)^2}{\Gamma_1p}\(H^{lm} - \xi^{lm}\)\;.
\end{equation}
For the sake of notational simplicity, we will from now on omit the
indices $l$ and $m$ for the perturbation variables. With the above
expansion, the evolution equations for $h_{ij}$ read:
\begin{eqnarray}
  \(\d_t + \I m\omega\)S_3Y_{lm} &=& \(2S_2' + 2\lam'S_2 - K_1\)Y_{lm}\non\\
  &&{} + 2\omega e^{-2\lam}\(V_3' - \lam'V_3\)\d_\phi Y_{lm}
  + 2\omega e^{-2\lam}\(V_4' - \lam'V_4\)\sin\theta\d_\theta Y_{lm}\;,\\
  \label{v34a}
  \d_t\(V_3\d_\theta - V_4\frac{\d_\phi}{\sin\theta}\)Y_{lm} &=&
  \(V_1' - \frac{2}{r}V_1 + e^{2\lam}\(S_2 - K_2\)\)\d_\theta Y_{lm}
  - \(V_2' - \frac{2}{r}V_2 - e^{2\lam}K_3\)
  \frac{\d_\phi Y_{lm}}{\sin\theta}\non\\
  &&{}- \omega \Lambda V_4\sin\theta Y_{lm}\;,\\
  \label{v34b}
  \d_t\(V_3 \d_\phi + V_4\sin\theta\d_\theta\)Y_{lm} &=&
  \(V_1' - \frac{2}{r}V_1 + e^{2\lam}\(S_2 - K_2\)\)\d_\phi Y_{lm}
  + \(V_2' - \frac{2}{r}V_2 - e^{2\lam}K_3\)\sin\theta\d_\theta Y_{lm}\;,\\
  \label{t123a}
  0 &=& \(2S_2 - \frac{\Lambda}{r}V_1 - K_4 + \frac{\Lambda}{r}K_5\)Y_{lm}
  + 2\omega e^{-2\lam}\(V_3\d_\phi Y_{lm} + V_4\sin\theta\d_\theta
  Y_{lm}\)\;,\\
  \label{t123b}
  0 &=& \(V_1 - K_5\) W_{lm}
  + \(V_2 - K_6\)\frac{X_{lm}}{\sin\theta}\;,\\
  \label{t123c}
  0 &=& \(V_1 - K_5\) X_{lm}
  - \(V_2 - K_6\)\sin\theta W_{lm}\;.
\end{eqnarray}
Still, in every equation a sum over all $l$ and $m$ is implied. From
Eqs.~(\ref{t123b}) and (\ref{t123c}) we immediately obtain our condition
for the shift components
\begin{eqnarray}
  V_1 &=& K_5\;,\\
  V_2 &=& K_6\;,
\end{eqnarray}
and from Eq.~(\ref{t123a}) it follows after multiplication with
$Y^*_{lm}$ and integration over the 2-sphere that
\begin{eqnarray}\label{s2}
  S_2 &=& \half K_4 - \omega e^{-2\lam}\(\I mV_3 + \cLa V_4\)\;,
\end{eqnarray}
where we have defined the operator $\cLa$, which couples the equations
of order $l$ to the equations of order $l+1$ and $l-1$, as
\begin{eqnarray}
  \cLa A_{lm} &:=& \sum_{l'm'}A_{l'm'}
  \int_{S_2}{Y^*_{lm}\sin\theta\d_\theta Y_{l'm'}d\Omega}
  \;=\; (l-1)Q_{lm}A_{l-1m} - (l+2)Q_{l+1m}A_{l+1m}\;,
\end{eqnarray}
with
\begin{eqnarray}
  Q_{lm} &:=& \sqrt{\frac{(l-m)(l+m)}{(2l-1)(2l+1)}}\;.
\end{eqnarray}
Later, we will also need
\begin{eqnarray}
  \cLb A_{lm} &:=& \sum_{l'm'}A_{l'm'}
  \int_{S_2}{\d_\theta Y^*_{lm}\sin\theta Y_{l'm'}d\Omega}
  \;=\; -(l+1)Q_{lm}A_{l-1m} + lQ_{l+1m}A_{l+1m}
\end{eqnarray}
and
\begin{eqnarray}
  \cLc A_{lm} &:=& \sum_{l'm'}A_{l'm'}
  \(l'(l'+1)\int_{S_2}{Y^*_{lm}\cos\theta Y_{l'm'}d\Omega}
    + \int_{S_2}{Y^*_{lm}\sin\theta\df{}{\theta}Y_{l'm'}d\Omega}\)\non\\
  &=& (l-1)(l+1)Q_{lm}A_{l-1m} + l(l+2)Q_{l+1m}A_{l+1m}\;.
\end{eqnarray}
The operator $\cLc$ can actually be expressed in terms of $\cLa$ and $\cLb$:
\begin{eqnarray}
  \cLc &=& -\half\(\cLa(\Lambda-2) + \cLb\Lambda\)\;.
\end{eqnarray}
By making use of these relations we can eliminate the spherical
harmonics and obtain the following simple set of evolution equations
for the metric perturbations:
\begin{eqnarray}
  \label{dts3}
    \(\d_t + \I m\omega\)S_3 &=& K_4' - K_1 + \lam'K_4
    - 2\omega'e^{-2\lam}\(\I mV_3 + \cLa V_4\)\;,\\
  \label{dtv3}
  \(\d_t + \I m\omega\)V_3 &=& K_5' - e^{2\lam}K_2
  - \frac{2}{r}K_5 + \half e^{2\lam}K_4\;,\\
  \label{dtv4}
  \(\d_t + \I m\omega\)V_4 &=& K_6' - e^{2\lam}K_3 - \frac{2}{r}K_6\;.
\end{eqnarray}
In a similar way, we obtain the evolution equations for the six
extrinsic curvature components, which are a little more lengthy:
\begin{eqnarray}
  \label{dtk1}
  \(\d_t + \I m\omega\)K_1 &=& e^{2\nu-2\lam}\bigg[\(\nu' + \frac{2}{r}\)S_3'
  - 2\frac{\Lambda}{r^2}V_3' + 2\lam'\frac{\Lambda}{r^2}V_3
  + 2\(\frac{\nu'}{r} - \frac{\lam'}{r} - \frac{e^{2\lam} - 1}{r^2}
  + e^{2\lam}\frac{\Lambda}{2r^2}\)S_3\bigg]\non\\
  &&{}+ 8\pi e^{2\nu}\(p + \eps\)C_s^{-2}\left[\(C_s^2 - 1\)H + \xi\right]
  - 2e^{-2\lam}\omega'\bigg[\I m\(K_5' - \frac{2}{r}K_5\)
  + \cLa\(K_6' - \frac{2}{r}K_6\)\bigg]\;,\\
  \label{dtk2}
  \(\d_t + \I m\omega\)K_2 &=&
  e^{2\nu-2\lam}\(\(\nu' + \frac{1}{r}\)S_3 - \frac{2}{r^2}V_3\)\non\\
  &&{} + \frac{\I mr^2}{2\Lambda}e^{-2\lam}\bigg[\omega'
  \(K_4' - K_1 + \lam'K_4 - 4\frac{\Lambda - 1}{r^2}K_5\)
  - 16\pi \varpi(p + \eps)
  \(e^{2\lam}K_4 + 2e^{2\nu}u_1\)\bigg]\non\\
  &&{} - \frac{\omega'e^{-2\lam}}{\Lambda}\cLa\(\(\Lambda - 2\)K_6\)\;,\\
  \label{dtk3}
  \(\d_t + \I m\omega\)K_3 &=& e^{2\nu-2\lam}\frac{\Lambda - 2}{r^2}V_4
  + e^{-2\lam}\frac{\omega'}{\Lambda}\(2\I mK_6 + \(\Lambda - 2\)\cLb K_5\)
  \non\\
  &&{}- \frac{r^2}{2\Lambda}e^{-2\lam}\cLb\bigg[
  \omega'\(K_4' - K_1 + \lam'K_4\)
  - 16\pi \varpi(p + \eps)\(e^{2\lam}K_4 + 2e^{2\nu}u_1\)\bigg]\;,\\
  \label{dtk4}
  \(\d_t + \I m\omega\)K_4 &=&
  e^{2\nu-2\lam}\bigg[S_3' + 2\(\nu' - \lam' + \frac{1}{r}\)S_3
  - \frac{2\Lambda}{r^2}V_3\bigg] + 8\pi r e^{2\nu}\(p + \eps\)C_s^{-2}
  \left[\(C_s^2 - 1\)H + \xi\right]\non\\
  &&{}+ r\(\cLa - \cLb\)\(\omega'K_3 + 16\pi e^{2\nu}\varpi(p + \eps)u_3\)\;,\\
  \label{dtk5}
  \(\d_t + \I m\omega\)K_5 &=&
  e^{2\nu-2\lam}\(V_3' + \(\nu' - \lam'\)V_3 - \half e^{2\lam}S_3\)\non\\
  &&{} + \frac{r^2}{\Lambda}\left\{\I m \left[\omega'\(\half K_4 - K_2\)
  - 16\pi e^{2\nu}\varpi(p + \eps)u_2\right]
  - \cLb\(\omega'K_3
  + 16\pi e^{2\nu}\varpi(p + \eps)u_3\)\right\}\;,\\
  \label{dtk6}
  \(\d_t + \I m\omega\)K_6 &=&
  e^{2\nu-2\lam}\(V_4' + \(\nu' - \lam'\)V_4\)\non\\
  &&{} - \frac{r^2}{\Lambda}\left\{\I m \(\omega'K_3
  + 16\pi e^{2\nu}\varpi(p + \eps)u_3\)
  + \cLb\left[\omega'\(\half K_4 - K_2\)
  - 16\pi e^{2\nu}\varpi(p + \eps)u_2\right]\right\}\;.
\end{eqnarray}
It is worthwhile to point out the symmetry between the polar and axial
equations. Each pair $V_3$ and $V_4$, $K_2$ and $K_3$, and $K_5$ and
$K_6$ represent the polar and axial counterparts of a metric or
extrinsic curvature perturbation. Thus, each associated pair of
equations (\ref{dtv3}) and (\ref{dtv4}), (\ref{dtk2}) and
(\ref{dtk3}), and (\ref{dtk5}) and (\ref{dtk6}) has basically the same
structure, with only the polar equations containing additional terms
as there are more polar variables than axial ones.

The last missing set of evolution equations is the one for the fluid
quantities, coming from $\delta\(T^{\mu\nu}_{\phantom{\mu\nu};\mu}\) =
0$ and from Eq.~(\ref{xir}):
\begin{eqnarray}
  \label{dth}
  \(\d_t + \I m\Omega\)H &=& C_s^2\bigg\{e^{2\nu-2\lam}\bigg[u_1' +
  \(2\nu' - \lam' + \frac{2}{r}\)u_1 - e^{2\lam}\frac{\Lambda}{r^2}u_2\bigg]
  + \half K_1 - \frac{\Lambda}{r^2}K_5 + \frac{1}{r}K_4\non\\
  &&{}\qquad+ \varpi e^{-2\lam}
  \bigg[\I m\(V_3' + \(\frac{2}{r} - \lam'\)V_3
  + e^{2\lam}\(H - \half S_3\)\)
  + \cLa\(V_4' + \(\frac{2}{r} - \lam'\)V_4\)\bigg]\bigg\}\non\\
  &&{}- \nu'\bigg[e^{2\nu-2\lam}u_1 + \half K_4
  + \varpi e^{-2\lam}\(\I mV_3 + \cLa V_4\)\bigg]\;,\\
  \label{dtu1}
  \(\d_t + \I m\Omega\)u_1 &=& H'
  + \frac{p'}{\Gamma_1 p}\bigg[\(\frac{\Gamma_1}{\Gamma} - 1\)H
  + \xi\bigg] -\I m\bigg[e^{-2\nu}\varpi\(K_5' - \frac{2}{r}K_5\)
  + \(\omega' + 2\varpi\(\nu' - \frac{1}{r}\)\)u_2\bigg]\non\\
  &&{}-\cLa\bigg[e^{-2\nu}\varpi\(K_6' - \frac{2}{r}K_6\)
  + \(\omega' + 2\varpi\(\nu' - \frac{1}{r}\)\)u_3\bigg]\;,\\
  \label{dtu2}
  \(\d_t + \I m\Omega\)u_2 &=& H
  + \frac{\varpi}{\Lambda}\(\I m\(2u_2 - e^{-2\nu}\(\Lambda - 2\)K_5\)
  + 2\cLc u_3 - e^{-2\nu}\cLa\(\(\Lambda - 2\)K_6\)\)
  - \frac{\I mr^2}{\Lambda}A\;,\\
  \label{dtu3}
  \(\d_t + \I m\Omega\)u_3 &=&
  2\frac{\varpi}{\Lambda}\bigg[\I m \(u_3  + e^{-2\nu}K_6\) - \cLc \(u_2
  + e^{-2\nu} K_5\)\bigg] + \frac{r^2}{\Lambda}\cLb A\;,\\
  \label{dtxi}
  \(\d_t + \I m\Omega\)\xi &=& \nu'\(\frac{\Gamma_1}{\Gamma} - 1\)
  \bigg[e^{2\nu-2\lam}u_1 + \half K_4
  + \varpi e^{-2\lam}\(\I mV_3 + \cLa V_4\)\bigg]\;,
\end{eqnarray}
where
\begin{eqnarray}
  A &=& \varpi C_s^2\bigg\{e^{-2\lam}\left[u_1'
    + \(2\nu' - \lam' + \frac{2}{r}\)u_1
    - e^{2\lam}\frac{\Lambda}{r^2}u_2\right]
  + e^{-2\nu}\left[\half K_1 - \frac{\Lambda}{r^2}K_5
    + \frac{1}{r}K_4\right]\bigg\}\non\\
  &&{}+ \left[\varpi\(\nu' - \frac{2}{r}\) + \omega'\right]
  \(e^{-2\lam}u_1 + \half e^{-2\nu}K_4\)\;.
\end{eqnarray}
In Eq.~(\ref{dth}), we have defined the sound speed $C_s$ as
\begin{eqnarray}
  C^2_s &=& \frac{\Gamma_1}{\Gamma}\frac{dp}{d\eps}\;.
\end{eqnarray}
The evolution equations comprise fourteen equations in total: four
axial and ten polar ones. In the non-rotating case, they are
equivalent to four wave equations, one for the axial and two for the
polar metric perturbations plus one wave equation for the fluid
variable $H$. The fluid equation for the axial velocity perturbation
$u_3$ vanishes in the non-rotating case, whereas equation (\ref{dtxi})
for the displacement variable $\xi$ does so in the barotropic case.

Finally we have the four constraint equations. The Hamiltonian
constraint reads
\begin{eqnarray}
  \label{HC}
  8\pi r^2e^{2\lam}\rho &=& rS_3' - \Lambda V_3'
  + \(1 - 2r\lam' + \half e^{2\lam}\Lambda\)S_3
  + \Lambda\(\lam' - \frac{1}{r}\)V_3\non\\
  &&{} + r^2e^{2\lam}
  \bigg[\I m\(\half\omega'e^{-2\nu}K_2 + 16\pi\varpi\(p + \eps\)u_2\)
  + \cLa\(\half\omega'e^{-2\nu}K_3 + 16\pi\varpi\(p + \eps\)u_3\)\bigg]\;,
\end{eqnarray}
and the three momentum constraints are
\begin{eqnarray}
  \label{MC1}
  8\pi r e^{2\nu}(p + \eps)u_1 &=&
  K_4' - \frac{\Lambda}{r}K_5' - K_1 + e^{2\lam}\frac{\Lambda}{2r}K_2
  + \frac{\Lambda}{r^2}\(1 + r\nu'\)K_5 - \nu'K_4
  + \frac{\I m}{4}r\omega'S_3\non\\
  &&{}-\(8\pi r\(p + \eps\)\varpi + 2e^{-2\lam}\omega'\)
  \(\I mV_3 + \cLa V_4\)\;,\\
  \label{MC2}
  16\pi r e^{2\nu}(p + \eps)u_2 &=&
  -rK_2' + rK_1 + \(r\nu' - r\lam' - 2\)K_2 - \frac{2}{r}K_5 + K_4
  + e^{-2\lam}\frac{r\omega'}{\Lambda}
  \(2\I m V_3- \(\Lambda - 2\)\cLb V_4\)\non\\
  &&{} + \frac{\I mr^3}{\Lambda}\bigg[\half e^{-2\lam}\omega'S_3'
  - 16\pi\varpi(p + \eps)\(S_3
  + C_s^{-2}\(\(C_s^2 + 1\)H - \xi\)\)\bigg]\;,\\
  \label{MC3}
  16\pi r e^{2\nu}(p + \eps)u_3 &=&
  -rK_3' + \(r\nu' - r\lam' - 2\)K_3 + \frac{\Lambda - 2}{r}K_6
  + e^{-2\lam}\frac{r\omega'}{\Lambda}
  \(2\I m V_4 + \(\Lambda - 2\)\cLb V_3\)\non\\
  &&{} - \frac{r^3}{\Lambda}\cLb\bigg[\half e^{-2\lam}\omega'S_3'
  - 16\pi\varpi(p + \eps)\(S_3 + C_s^{-2}\(\(C_s^2 + 1\)H - \xi\)\)\bigg]\;.
\end{eqnarray}
The axial equations (\ref{dtv4}), (\ref{dtk3}), (\ref{dtk6}),
(\ref{dtu3}) and (\ref{MC3}) without the coupling terms to the polar
perturbations are equivalent to Eqs.~(7)--(10) and (12) of Ruoff \&
Kokkotas (2001b). Note, however, that therein a slightly different
definition of the perturbation variables has been chosen.

\section{The non-rotating limit}

Although the non-rotating limit is well described by the wave
equations given by Allen et al.~(1998), it is instructive to consider
it in the BCL gauge. This is obtained by setting $\Omega$ and $\omega$
to zero in all the evolution equations (\ref{dts3})--(\ref{dtxi}) and
the constraints (\ref{HC})--(\ref{MC3}). As is well known, in this
case the polar and axial parts of the equations completely decouple.
For barotropic perturbations ($\Gamma_1 = \Gamma$), the polar
evolution equations can then be easily transformed into three wave
equations for the rescaled metric variables $S = e^{\nu-\lam} S_3$ and
$V = e^{\nu-\lam}V_3/r$ and the rescaled fluid variable $\tilde H =
e^{-\nu-\lam}H/r$:
\begin{eqnarray}
  \label{waveS}
  \dff{S}{t} &=& \dff{S}{r_*}+ e^{2\nu-2\lam}\bigg[
  \(\nu'\(\nu' - \lam'\) + 3\frac{\nu'}{r} + \frac{\lam'}{r} - \frac{3}{r^2}
  - e^{2\lam}\frac{\Lambda - 1}{r^2} - \lam''\)S
  + \frac{4\Lambda}{r^2}\(1 - r\nu'\)V\bigg]\non\\
  &&{} + 8\pi e^{2\nu - 2\lam}\bigg[\(C_s^2 - 1\)
  \(\tilde \rho' + \(\nu' - \frac{1}{r}\)\tilde \rho\)
  + \(C_s^2\)'\tilde \rho\bigg]\;,\\
  \label{waveV}
  \dff{V}{t} &=& \dff{V}{r_*}
  + e^{2\nu-2\lam}\bigg[\(\frac{\nu'}{r} - \frac{\lam'}{r}
  + 2\frac{e^{2\lam} - 1}{r^2} - e^{2\lam}\frac{\Lambda}{r^2}\)V
  - e^{2\lam}\(\frac{\nu'}{r} + \frac{\lam'}{r} - \frac{1}{r^2}\)S\bigg]\non\\
  &&{} + 4\pi e^{2\nu}\(C_s^2 - 1\)\tilde \rho\;,\\
  \label{waveH}
  \dff{\tilde H}{t} &=& e^{2\nu-2\lam}\bigg\{C_s^2\dff{\tilde H}{r}
  - \(C_s^2\lam' + \nu'\)\df{\tilde H}{r}\non\\
  &&{}\qquad\quad + \bigg[C_s^2\(\lam'\(\frac{3}{r} + \lam'\)
  + \frac{e^{2\lam} - 1}{r^2} - \lam'' - e^{2\lam}\frac{\Lambda}{r^2}\)
  + \frac{\lam'}{r} + 2\frac{\nu'}{r}
  + \frac{\nu'}{C_s^2}\(\nu' + \lam'\)\bigg]\tilde H\bigg\}\non\\
  &&{} + e^{2\nu}\bigg\{\frac{r\nu'}{2}\(C_s^2 - 1\)\df{S}{r}
  + \bigg[C_s^2\(\frac{\nu'}{2}\(r\lam' - r\nu' + 6\)
  + \lam' - \frac{e^{2\lam} - 1}{r}\)
  + \frac{\nu'}{2}\(r\lam' - r\nu' - 2\)\bigg]S\non\\
  &&{}\qquad\quad- \nu'\Lambda\(C_s^2 - 1\)V\bigg\}\;.
\end{eqnarray}
In Eqs.~(\ref{waveS}) and (\ref{waveV}), $r_*$ is the well-known
tortoise coordinate, which is related to $r$ through
\begin{equation}
  \frac{d}{dr_*} = e^{\nu-\lam}\frac{d}{dr}\;.
\end{equation}
Furthermore, one can express the energy density $\tilde\rho$ in terms
of $\tilde H$, which in the barotropic case reduces to
\begin{equation}
  \tilde \rho = \frac{p + \eps}{C_s^2}\tilde H\,.
\end{equation}
Although the equations in the first order form are quite simple, the
above set of wave equations is more complicated than the equivalent
set in the Regge--Wheeler gauge (Eqs.~(14), (15) and (16) of Allen et
al.~1998). This is particular so for the way in which the fluid
variable $\tilde\rho$ (or equivalently $\tilde H$) couples to the
metric variable $S$, where the derivative of $\tilde\rho$ enters. If
the stellar model is based on a polytropic equation of state $p =
\kappa \eps^\Gamma$, then the behaviour of $\tilde\rho$ at the stellar
surface strongly depends on the polytropic index $\Gamma$. As
discussed in Ruoff (2001a), $\tilde\rho$ actually diverges for $\Gamma
> 2$. In this case the metric quantity $S$ would not even be ${\cal
  C}^0$, which can be troublesome for the numerical convergence.
Although this could be a drawback of the BCL gauge, there is a clear
advantage if one is interested in computing the gauge invariant
Zerilli function $Z$ in the exterior.  Following Moncrief (1974), the
definition of the Zerilli function is
\begin{eqnarray}
  Z &=& \frac{r^2\(\Lambda k_1 + 4e^{-4\lam}k_2\)}{r\(\Lambda - 2\) + 2M}
\end{eqnarray}
with
\begin{eqnarray}
  k_1 = -2e^{-\lam-\nu}V
\end{eqnarray}
and
\begin{eqnarray}
  k_2 = \half e^{3\lam-\nu}S\;.
\end{eqnarray}
In terms of $S$ and $V$ this gives us
\begin{eqnarray}\label{Zer}
  Z &=& \frac{2r^2e^{-\lam-\nu}}{r\(\Lambda - 2\) + 2M}\(S - \Lambda V\)\;,
\end{eqnarray}
which is a simple algebraic relation in contrast to the relation in
the Regge--Wheeler gauge, which includes a spatial derivative of one
of the metric perturbations (see Eq.~(20) of Allen et al.~1998, or
Eq.~(60) of Ruoff 2001a). In the Regge--Wheeler gauge, the two metric
variables ($S$ and $F$ in the notation of Allen et al.~1998, and $S$
and $T$ in the notation of Ruoff 2001a) have different asymptotic
behaviour at infinity, in particular one ($F$ or $T$) is linearly
growing with $r$.  It is only through the delicate cancellation of the
growing terms that the Zerilli function remains finite at infinity.
However, this cancellation can only happen if both metric variables
exactly satisfy the Hamiltonian constraint.  Any (numerical) violation
leads to an incomplete cancellation, and the Zerilli function starts
to grow at large radii. This makes it very difficult in the numerical
time evolution to extract the correct amount of gravitational
radiation emitted from the neutron star. With the above relation
(\ref{Zer}), we do not expect such difficulties to occur.

\section{Conclusions}

We have presented the derivation of the perturbation equations for
slowly rotating relativistic stars using the BCL gauge, which has been
first used by Battiston, Cazzola \& Lucaroni in 1971.  This gauge is
defined by setting $\alpha$, $h_{\theta\theta}$, $h_{\theta\phi}$ and
$h_{\phi\phi}$ to zero. In the non-rotating case, the vanishing shift
condition leads to a complete vanishing of $h_{tt}$, however, in the
rotating case $h_{tt}$ becomes non-zero (see also Appendix).  The
advantage of the BCL gauge over the Regge--Wheeler gauge is that in
the ADM formalism, the evolution equations a priori do not contain any
second order spatial derivatives. Instead, one is immediately lead to
a hyperbolic set of first order evolution equations, which can be
directly used for the numerical time evolution without many further
manipulations. Although it is in principle also possible to derive a
hyperbolic set in the Regge--Wheeler gauge, the procedure is rather
tedious and requires the introduction of carefully chosen new
variables in order to replace the second order or mixed derivatives.

The perturbation equations for slowly rotating relativistic
stars form a set of fourteen evolution equations plus four
constraints. In the non-rotating barotropic case, it is possible to
cast the equations into a system of four wave equations, three for the
polar and axial metric perturbations and one for the polar fluid
perturbation, as it is the case in the Regge--Wheeler gauge. Although
these wave equations are not simpler than the corresponding ones in
the Regge--Wheeler gauge, the first order system actually is. Maybe the
main advantage is the simple algebraic relation of the metric
variables to the Zerilli function. It was demonstrated by Ruoff
(2001a) that the accurate numerical evaluation of the Zerilli function
in the Regge--Wheeler gauge is somewhat difficult and requires high
resolution because a small numerical violation of the Hamiltonian
constraint can lead to very large errors in the Zerilli function. This
should not be the case in the BCL gauge as the relation (\ref{Zer})
does not involve any derivatives.

A further advantage of these evolution equations is that the inclusion
of the source terms describing a particle orbiting the star can be
very easily accomplished. This is not the case for the Regge--Wheeler
gauge, as even for the non-rotating case, one is forced to include
second order derivatives of the source terms (Ruoff 2001b). Since the
source terms contain $\delta$-functions, one has to deal with second
order derivatives thereof. In the axial case, no derivatives appear,
and the perturbation equations with the source terms, which are given
in Ruoff (2001b), are very simple. We expect the same to be the case
for the polar equations in the BCL gauge. Here, it should be possible
to plug the source terms into the equations for the extrinsic
curvature without getting any derivatives.

In subsequent papers, we will present results from the numerical
evolution of the perturbation equations of slowly rotating
relativistic stars in the BCL gauge with the particular focus on
oscillations modes which are unstable with respect to gravitational
radiation. As a step further, we will include the contribution of a
test particle acting as a source of excitation for the stellar
oscillations.

\section*{acknowledgments}
We thank Nils Andersson, Luciano Rezzolla and Nikolaos Stergioulas for
many helpful comments. J.R. is supported by the Marie Curie Fellowship
No.  HPMF-CT-1999-00364. A.S. is supported by the Greek National
Scholarship foundation (I.K.Y.). This work has been supported by the
EU Programme 'Improving the Human Research Potential and the
Socio-Economic Knowledge Base' (Research Training Network Contract
HPRN-CT-2000-00137).

\section*{Appendix: The perturbation equations following from
Einstein's equations}

Kojima (1992) derived the perturbation equations in the Regge--Wheeler
gauge directly from the linearized Einstein equations without
resorting to the ADM formalism. In this section we repeat this
calculation using the BCL gauge. In order to facilitate the comparison
with Kojima's equations, who uses the more familiar notation of Regge
\& Wheeler (1957), we switch to a similar notation.  In the
Regge--Wheeler gauge, the quantities $h_0$ and $h_1$ denote the axial
perturbations of $h_{t\{\phi,\theta\}}$ and $h_{r\{\phi,\theta\}}$,
respectively, whereas the corresponding polar perturbations are set to
zero. Since in the BCL gauge, the latter do not vanish, we denote them
by $h_{0,p}$ and $h_{1,p}$, respectively, and to avoid confusion we
denote the axial ones by $h_{0,a}$ and $h_{1,a}$. The remaining
non-zero polar perturbations are then $H_1$ and $H_2$. Thus, the
expansion of the metric in BCL gauge reads in this notation:
\begin{equation}
  h_{\mu\nu} = \sum_{lm} \(
  \begin{array}{cccc}
    -2 \omega \(h^{lm}_{0,p} \d_\phi + h^{lm}_{0,a} \sin\theta \d_\theta\)
    & H^{lm}_1 &
    h^{lm}_{0,p} \d_\theta - h^{lm}_{0,a}/\sin\theta\d_\phi &
    h^{lm}_{0,p} \d_\phi + h^{lm}_{0,a} \sin\theta \d_\theta\\
    \star & e^{2\lambda}H^{lm}_2 &
    h^{lm}_{1,p} \d_\theta - h^{lm}_{1,a}/\sin\theta\d_\phi &
    h^{lm}_{1,p}\d_\phi + h^{lm}_{1,a}\sin\theta \d_\theta\\
    \star & \star & 0 & 0 \\
    \star & \star & 0 & 0 \\
  \end{array}\)Y_{lm}\;.
  \label{hlm}
\end{equation}
Note that here $h_{tt}$ is not zero, which is a consequence of the
relation between the perturbation of the lapse $\alpha$ and $h_{tt}$
given by Eq.~(\ref{ah_rel}). The relation between the above variables
and the ones used in the previous sections is the following (we again
omit the indices $l$ and $m$):
\begin{eqnarray}
  H_1 &=& e^{2\lambda}K_4 - \omega \( \I m V_3 + \cLa V_4 \)\;,\\
  H_2 &=& S_3\;,\\
  h_{0,p} &=& K_5\;,\\
  h_{0,a} &=& K_6\;,\\
  h_{1,p} &=& V_3\;,\\
  h_{1,a} &=& V_4\;,\\
  R &=& -u_1\;,\\
  V &=& -u_2\;,\\
  U &=& -u_3\;.
\end{eqnarray}
The extrinsic curvature components can be expressed as
\begin{eqnarray}
  K_1 &=& 2 e^{-2\lambda} \left\{H_1' -\lambda'H_1
    + \omega \left[ \I m\(h_{1,p}' - \lambda'h_{1,p}\)
      + \cLa \(h_{1,a}' - \lambda' h_{1,a}\)\right]\right\}
  - \dot{H}_2 - \I m \omega H_2\;,\\
  K_2 &=& e^{-2\lambda} \left[ h_{0,p}' - {2 \over r} h_{0,p}
    +  H_1 - \dot{h}_{1,p} + \omega  \cLa h_{1,a}\right]\;,\\
  K_3 &=& e^{-2\lambda} \left[ h_{0,a}' - {2 \over r} h_{0,a} + \dot{h}_{1,a}
    - \I m \omega h_{1,a} \right]\;.
\end{eqnarray}
A very often occurring combination of variables in the perturbation
equations is $h_0' - \dot{h}_1$ for both the axial and polar cases,
which we abbreviate with the following functions
\begin{eqnarray}
  Z_a &=& h_{0,a}' - \dot{h}_{1,a}\;,\\
  \label{Zeta}
  Z_p &=& h_{0,p}' - \dot{h}_{1,p}\;.
  \label{Zetae}
\end{eqnarray}
The equations coming from the $(tt)$, $(tr)$, $(rr)$ and the addition
of the $(\theta\theta)$ and $(\phi\phi)$ components can be written as
\begin{equation}
  A^{(I)}_{lm} + \I m C^{(I)}_{lm} + \cLb B^{(I)}_{lm}
  + \cLd \tilde{A}^{(I)}_{lm} = 0\;,\\
\label{tt}
\end{equation}
with
\begin{eqnarray}
  \cLd A_{lm} &:=& -\half\(\cLa + \cLb\) A_{lm}
  \;=\; Q_{lm}A_{l-1m} + Q_{l+1m}A_{l+1m}
\end{eqnarray}
and
%%%%%%%%%%%%%%%%%%%%%%%%%%%%%%  tt  %%%%%%%%%%%%%%%%%%%%%%%%%%%%%%%%%%
\begin{eqnarray}
  A^{(tt)} &=& {2 e^{2\nu} \over r^2}
  \left[ r H_2' - \Lambda h_{1,p}'
    -16 \pi r^2 e^{2\lambda} C_s^{-2} \( H  - \xi \)
    + \Lambda \( \lambda'  - {1 \over r}  \) h_{1,p}
    + \( 1 - 2r \lambda' + {  \Lambda e^{2\lambda} \over 2 } \) H_2
  \right]\;,\\
  \label{A1tt}
  \tilde{A}^{(tt)} &=& 0\;, \\
  \label{tildeA1}
  B^{(tt)} &=& 2 \omega  Z_a'
  + \( \omega' - 2\omega \( \lambda' + \nu' - {2 \over r} \) \) Z_a
  - {4 \omega  \over r}  h_{0,a}'
  - 32 \pi \Omega \(p+\eps\) e^{2\nu+2\lambda}  U\non \\
  &&{}+ {2 \over r} \left[ -\omega'
    + \omega \( 2\nu' + 2\lambda' - { 2 \over r }
    - e^{2\lambda} {\Lambda - 2 \over r} \)
  \right] h_{0,a}\;, \\
  \label{B1tt}
  C^{(tt)} &=& 2 \omega
  \( Z_p' - H_1' + e^{2\lambda} \dot{H}_2 \)
  + \( \omega' - 2\omega \( \lambda' + \nu'\) \) h_{0,p}'
  - {2\over r}\left[\omega'- 2\omega\(\nu' + \lambda'
    - {e^{2\lambda}-1 \over r}\)\right] h_{0,p} \non \\
  &&{}- \left[\omega' - 2\omega\(\nu' + \lambda' -{2 \over r}\)\right]
  \dot{h}_{1,p} + \( \omega' + 2\omega( \lambda' - \nu') \) H_1
  - 32\pi \Omega e^{2\nu+2\lambda} \(p+\eps\) V\;,\\
  \label{C1tt}
%%%%%%%%%%%%%%%%%%%%%%%%%%%%%%%%%  tr  %%%%%%%%%%%%%%%%%%%%%%%%%%%%%%%%%
  A^{(tr)} &=&
  {2 \over r} \dot{H}_2 + {\Lambda \over r^2}
  \( Z_p + H_1 - 2 h_{0,p}' + 2\nu' h_{0,p} \)
  + 16 \pi\(p + \eps\)\(e^{2\nu}R - H_1\)\;,\\
  \label{A2tr}
  \tilde{A}^{(tr)} &=& {2 \Lambda \omega \over r^2} h_{1,a} \;,\\
  \label{tildeA2}
  B^{(tr)} &=&  \left[ {\Lambda \omega \over r^2}
    - 16 \pi  \Omega \(p + \eps\) \right] h_{1,a} \;,\\
  \label{B2tr}
  C^{(tr)} &=&
  \( {2 \omega \over r}  + {\omega' \over 2} \) H_2
  - 16 \pi \Omega \(p +\eps\) h_{1,p} \;,\\
  \label{C2tr}
%%%%%%%%%%%%%%%%%%%%%%%%%%%%%%%  rr  =
%%%%%%%%%%%%%%%%%%%%%%%%%%%%%%%%%%%%
  A^{(rr)} &=& \dot{H}_{1}
  + e^{2\nu}{\Lambda\over 2r}\( \lambda' + {1 \over r} \) h_{1,p}
  - 4 \pi r e^{2\nu+2\lambda} \(p+\eps\) H
  - {e^{2\nu} \over 2r} \left[ \( 2r \nu' + 1 \)
    - {\Lambda \over 2} e^{2\lambda} \right] H_2
  - e^{2\nu}{\Lambda \over 2r} h_{1,p}' \;,\\
  \label{A3rr}
  \tilde{A}^{(rr)} &=& 0 \;,\\
  \label{tildeArr}
  {B}^{(rr)} &=& \omega  h_{0,a}'
  + {\omega' \over 2} h_{0,a}
  - \( \omega + {r \omega' \over 4}  \) Z_a \;,\\
  \label{B3rr}
  C^{(rr)} &=& \omega h_{0,p}'
  - \( \omega  +  {r \omega' \over 4} \) Z_p
  +  \( \frac{\omega'}{2} - e^{2\lambda}{\Lambda \omega \over r }
  \) h_{0,p}
  +  \( \omega + {r \omega' \over 4} \) H_1 \;,\\
  \label{C3rr}
%%%%%%%%%%%%%%%%%%%%%%%%%%%%%%%  theta + phi  %%%%%%%%%%%%%%%%%%%%%%%%%%%%%%%
  A^{(\theta\theta+\phi\phi)} &=& -\ddot{H}_2
  + 2 e^{-2\lambda} \left[\dot{H}_1'
    + \( {1 \over r} - \lambda'\)\dot{H}_1\right]
  -  e^{2\nu-2\lambda} \( \nu' + {1 \over r} \) H_2'
  - {\Lambda \over r^2 }\( \dot{h}_{0,p} - e^{2\nu-2\lambda} h_{1,p}'\)\non\\
  &&{}- 16 \pi e^{2\nu} (p+\eps) H
  - e^{2\nu} \( { \Lambda \over 2r^2 } + 16 \pi p \) H_2
  + { \Lambda \over r^2 }  e^{2\nu-2\lambda} \( \nu' -\lambda' \)h_{1,p}\;, \\
  \label{A4thph}
  \tilde{A}^{(\theta\theta+\phi\phi)} &=& 0 \;,\\
  \label{tildeA4thph}
  {B}^{(\theta\theta+\phi\phi)} &=&
  2 \omega e^{-2\lambda} \( {h}_{0,a}'' - Z_a'
  + \({1\over r} - \lambda'\)\(h_{0,a} - Z_a\)\)
  + 2\omega'e^{-2\lambda}\(h_{0,a}' - {2\over r}h_{0,a}\)
  - 16 \pi e^{2\nu} \varpi (p+\eps) U \;,\\
  \label{B4thph}
  {C}^{(\theta\theta+\phi\phi)} &=&
  2 \omega e^{-2\lambda}\left[ h_{0,p}'' - Z_p' + H_1'
    - e^{2\lambda}\(\dot{H}_2 + {\Lambda \over r^2} h_{0,p}\)
  + \({1\over r} - \lambda'\)\(\dot{h}_{1,p} + H_1\)\right]\non \\
  &&{}+ e^{-2\lambda} \omega'\(H_1 + 2h_{0,p}' - {4 \over r}h_{0,p}\)
  - 16\pi\varpi e^{2\nu}\(p + \eps\) V\;.
  \label{C4thph}
\end{eqnarray}
%%%%%%%%%%%%%%%%%%%%%%%%%%%%%%%%%%%%%%%%%%%%%%%%%%%%%%%%%%%%%%%%%%%%%%%%%%%%%
The $(t\theta)$ and $(r\theta)$ components are
\begin{equation}
  \Lambda a^{(I)}_{lm} + \I m d^{(I)}_{lm} + \cLc \tilde{a}^{(I)}_{lm}
  + \cLb \eta^{(I)}_{lm} = 0\;,
  \label{even}
\end{equation}
with
%%%%%%%%%%%%%%%%%%%%%%%%%%%%%%%%%%%%%%%%%%%%%%%%%%%%%%%%%%%%%%%%%%%%%%%%%%%%%
%%%%%%%%%%%%%%%%%%%%%%%%%%%%    t theta   %%%%%%%%%%%%%%%%%%%%%%%%%%%%%%%%%%%
%%%%%%%%%%%%%%%%%%%%%%%%%%%%%%%%%%%%%%%%%%%%%%%%%%%%%%%%%%%%%%%%%%%%%%%%%%%%%
\begin{eqnarray}
  {a}^{(t\theta)} &=&
  - \dot{H}_2 + e^{-2\lambda} \left[H_1' - Z_p' + {2 \over r} h_{0,p}'
    + \(\lambda' + \nu' - {2\over r} \)Z_p + \(\nu' - \lambda'\)H_1
    - { 2 \over r^2}\(r\lambda' - r\nu' + e^{2\lambda} - 1\)h_{0,p}
  \right]\non\\
  &&{} + 16 \pi e^{2\nu}\(p+\eps\)V\;,\\
  {d}^{(t\theta)} &=&  e^{-2\lambda}\left[
  2\Lambda \omega \(h_{1,p}' +  \( {1 \over r} - \nu' \)h_{1,p}\)
  + \omega' \({r^2\over 2} H_2' + 2h_{1,p}\)\right]
  - 16\pi r^2 \varpi \(p+\eps\)\(H_2 + \(1 + C_s^{-2}\)H + C_s^{-2}\xi\)\;,\\
  \label{dttheta}
  \tilde{a}^{(t\theta)} &=& 2 \omega' e^{-2\lambda} h_{1,a} \;,\\
  \label{tildea1ttheta}
  \eta^{(t\theta)} &=& -\Lambda \omega e^{-2\lambda}
  \left[ h_{1,a}' + \(\nu' - \lambda'\) h_{1,a} \right]\;, \\
  \label{etattheta}
%%%%%%%%%%%%%%%%%%%%%%%%%%%%%%%%%%%%%%%%%%%%%%%%%%%%%%%%%%%%%%%%%%%%%%%%
%%%%%%%%%%%%%%%%%%%%%%%%%%%%%%%%% r theta %%%%%%%%%%%%%%%%%%%%%%%%%%%%%%
%%%%%%%%%%%%%%%%%%%%%%%%%%%%%%%%%%%%%%%%%%%%%%%%%%%%%%%%%%%%%%%%%%%%%%%%
  a^{(r\theta)} &=& - \dot{Z}_p
  + {1 \over r}
  \( 2 e^{2\nu-2\lambda} - {\Lambda \over 2} \) h_{1,p}'
  +\left[8\pi e^{2\nu}\(p+\eps\)
    + {\Lambda \over 2r^2} \( 1 + r \lambda' \)
    -{2\over r^2} e^{2\nu}
  \right] h_{1,p}\non \\
  &&{}+ \left[ \nu' \( e^{2\nu} - 1 \)
    + {1 \over 2r} \( {\Lambda\over 2}e^{2\lambda} - 1 \)\right] H_2
  - 4 \pi r e^{2\lambda} \(p+\eps\) H\;, \\
  \label{artheta}
  d^{(r\theta)} &=& 16 \pi r^2 \varpi\(p+\eps\)\(H_1 + e^{2\nu} R\)
  - \omega \Lambda\(H_1 + Z_p -  h_{0,p}\)
  +  \omega'\({r^2\over 2}\dot{H}_2 - 2\(\Lambda+2\)h_{0,p}\)\;,\\
  \label{dtth}
  \tilde{a}^{(r\theta)} &=& 2 \omega' h_{0,a}\;,\\
  \label{tildeartheta}
  \eta^{(r\theta)} &=&
  \Lambda \left[ \omega \( h_{0,a}' - Z_a \) + \omega' h_{0,a} \right]\;.
  \label{etartheta}
\end{eqnarray}
%%%%%%%%%%%%%%%%%%%%%%%%%%%%%%%%%%%%%%%%%%%%%%%%%%%%%%%%%%%%%%%%%%%%%%%%
From the $(t\phi)$ and $(r\phi)$ components we get
\begin{equation}
  \Lambda b^{(I)}_{lm} + \I m c^{(I)}_{lm} + \cLc \tilde{b}^{(I)}_{lm}
  + \cLb \zeta^{(I)}_{lm} = 0\;,
  \label{odd}
\end{equation}
with
%%%%%%%%%%%%%%%%%%%%%%%%%%%%%%%%%%%%%%%%%%%%%%%%%%%%%%%%%%%%%%%%%%%%%%%%
%%%%%%%%%%%%%%%%%%%%%%%%% t phi %%%%%%%%%%%%%%%%%%%%%%%%%%%%%%%%%%%%%%%%
%%%%%%%%%%%%%%%%%%%%%%%%%%%%%%%%%%%%%%%%%%%%%%%%%%%%%%%%%%%%%%%%%%%%%%%%
\begin{eqnarray}
  b^{(t\phi)} &=& - Z_a'
  + \( \nu' + \lambda' - {2 \over r} \) Z_a
  + {2 \over r}  h_{0,a}'
  + 16 \pi  e^{2\nu+2\lambda} \(p+\eps\) U
  - \left[ {2 \over r}\( \nu' + \lambda' - {1 \over r} \)
    - e^{2\lambda} {\Lambda - 2  \over r^2} \right] h_{0,a} \;,\\
  \label{btphi}
  c^{(t\phi)} &=& -3 \Lambda \omega  h_{1,a}'
  +\left[ \Lambda \omega \( 3 \lambda' - \nu' -{2 \over r} \)
    - \(\Lambda - 2\) \omega' \right] h_{1,a} \;,\\
  \label{ctphi}
  \tilde{b}^{(t\phi)} &=& - 2 e^{-2\lambda} \omega' h_{1,p} \;,\\
  \label{btildetphi}
  \zeta^{(t\phi)} &=&
  2\omega\Lambda\(e^{2\lambda}H_2 + \(\nu' -\lambda'\)h_{1,p}\)
  + \omega'\(\Lambda h_{1,p} - {r^2\over 2}H_2'\)
  + 16 \pi r^2 e^{2\lambda} \varpi \(p+\eps\)
  \left[ H_2 + \( 1 + C_s^{-2} \) H + C_s^{-2} \xi \right]\;,\\
  \label{dtphi}
%%%%%%%%%%%%%%%%%%%%%%%%%%%%%%%%%%%%%%%%%%%%%%%%%%%%%%%%%%%%%%%%%%%%%%%%
%%%%%%%%%%%%%%%%%%%%%%%%% r phi  %%%%%%%%%%%%%%%%%%%%%%%%%%%%%%%%%%%%%%%
%%%%%%%%%%%%%%%%%%%%%%%%%%%%%%%%%%%%%%%%%%%%%%%%%%%%%%%%%%%%%%%%%%%%%%%%
  b^{(r\phi)} &=&
  \dot{Z}_a - {2 \over r} e^{2\nu-2\lambda}{h}_{1,a}'
  - e^{2\nu} \left[ { \Lambda - 2 \over r^2 } + {2 \over r} e^{-2\lambda}
    \( \nu' - \lambda' \)\right] h_{1,a} \;,\\
  \label{brtheta}
  c^{(r\phi)} &=& \Lambda \omega h_{0,a}'
  + 2\left[\(\Lambda + 1\)\omega' - {\Lambda\omega\over r}\right] h_{0,a}\;,\\
  \label{crph}
  \tilde{b}^{(r\phi)} &=& 2 \omega' h_{0,p}\;, \\
  \label{tildebrphi}
  \zeta^{(r\phi)} &=& \omega'\(\Lambda h_{0,p} - {r^2\over 2}\dot{H}_2\)
  - 16\pi r^2\varpi \(p+\eps\)\(e^{2\nu}R + H_1\)\;.
  \label{drphi}
\end{eqnarray}
%%%%%%%%%%%%%%%%%%%%%%%%%%%%%%%%%%%%%%%%%%%%%%%%%%%%%%%%%%%%%%%%%%%%%%%
%%%%%%%%%%%%%%%%%%%%%%%%%%% theta phi %%%%%%%%%%%%%%%%%%%%%%%%%%%%%%%%%%
%%%%%%%%%%%%%%%%%%%%%%%%%%%%%%%%%%%%%%%%%%%%%%%%%%%%%%%%%%%%%%%%%%%%%%%%
From the $(\theta\phi)$ and the subtraction of $(\theta\theta)$ and
$(\phi\phi)$ components we get
\begin{eqnarray}
  \Lambda s_{lm} - \I m f_{lm} + \cLb g_{lm} &=& 0\;,\\
  \label{thph}
  \Lambda t_{lm} + \I m g_{lm} + \cLb f_{lm} &=& 0\;,
  \label{thminusph}
\end{eqnarray}
with
\begin{eqnarray}
  f &=& \omega' r^2 e^{-2\lambda}
  \( Z_p - {2 \over r} h_{0,p}\)
  - 16 \pi r^2 \varpi e^{2\nu} \(p+\eps\) V \;,\\
  \label{fthhp}
  g &=& -\omega' r^2 e^{-2\lambda}
  \( Z_a - { 2 \over r} h_{0,a} \)
  + 16 \pi r^2 \varpi e^{2\nu} \(p+\eps\) U \;,\\
  \label{gthph}
  s &=& - \dot{h}_{0,p}
  +  e^{2\nu-2\lambda} \(h_{1,p}' + \(\nu' - \lambda'\) h_{1,p}\)
  - {e^{2\nu} \over 2}H_2 - \I m \omega h_{0,p} \;,\\
  \label{sthph}
  t &=& -\dot{h}_{0,a}
  + e^{2\nu-2\lambda} \(h_{1,a}' + \(\nu' - \lambda'\)h_{1,a}\)
  - \I m \omega  h_{0,a}\;.
  \label{tthhp}
\end{eqnarray}
These equations are fully equivalent to the ones derived within the
ADM formalism. Although they still contain some second order
derivatives, they can be easily brought by introduction of a few
auxiliary variables into a first order hyperbolic form or even into
characteristic form, which is very useful for the numerical evolution.

\label{lastpage}
\end{document}